\documentclass[apj]{emulateapj}
\submitted{ApJ, accepted 21/11/2012}

\usepackage{amsmath}
\usepackage{amsfonts}
\usepackage{amsthm}
\usepackage{amssymb}
\usepackage{graphicx}
\usepackage{mathrsfs}
\usepackage{latexsym}
\usepackage{graphics}
\usepackage{cases}
\usepackage{hyperref}
\usepackage{mathtools}
\usepackage{dsfont}
\usepackage{enumerate}
\usepackage{ulem}
\newcommand{\be}{\begin{equation}}
\newcommand{\ee}{\end{equation}}
\newcommand{\bea}{\begin{eqnarray}}
\newcommand{\eea}{\end{eqnarray}}
\newcommand{\bes}{\begin{equation}\begin{split}}
\newcommand{\ees}{{\end{split}\end{equation}}}
\newcommand{\eq}[1]{eq.~(\ref{#1})}

\renewcommand{\d}{{\rm d}}

\newcommand{\ensavg}[1]{\langle#1\rangle}
\newcommand{\avg}[1]{\overline{#1}}
\newcommand{\Abs}[1]{\left|#1\right|}
\newcommand{\abs}[1]{|#1|}

\newcommand{\h}{\ell}
\newcommand{\hr}{\h(r)}
\newcommand{\mdm}{m_{\rm DM}}
\renewcommand{\r}{\mathbf{r}}
\renewcommand{\k}{\mathbf{k}}
\newcommand{\q}{\mathbf{q}}
\newcommand{\s}{\mathbf{s}}
\renewcommand{\t}{\mathbf{t}}
\renewcommand{\a}{\mathbf{a}}
\renewcommand{\b}{\mathbf{b}}
\newcommand{\dr}{\delta(\r)}
\newcommand{\dk}{\hat{\delta}(\k)}
\newcommand{\er}{\epsilon(\r)}
\newcommand{\ek}{\hat{\epsilon}(\k)}
\newcommand{\R}{{\mathds{R}^D}}
\newcommand{\hmpc}{\,h^{-1}{\rm Mpc}}

\begin{document}

\title{A robust measure of cosmic structure beyond the power-spectrum:\\cosmic filaments and the temperature of dark matter}

\author{D. Obreschkow$^1$}
\author{C. Power$^1$}
\author{M. Bruderer$^2$}
\author{C. Bonvin$^{3,4}$}
\affiliation{$^1$~International Centre for Radio Astronomy Research (ICRAR), M468, University of Western Australia, 35 Stirling Hwy, Crawley, WA 6009, Australia}
\affiliation{$^2$~Institut f\"{u}r Theoretische Physik, Albert-Einstein Allee 11, Universit\"{a}t Ulm, 89069 Ulm, Germany}
\affiliation{$^3$~Kavli Institute for Cosmology Cambridge and Institute of Astronomy, Madingley Road, Cambridge CB3 OHA, UK}
\affiliation{$^4$~DAMTP, Centre for Mathematical Sciences, Wilberforce Road, Cambridge CB3 OWA, UK}


\date{\today}

\begin{abstract}
We discover that the mass of dark matter particles $\mdm$ is imprinted in phase-correlations of the cosmic density field more significantly than in the 2-point correlation. In particular, phase-correlations trace $\mdm$ out to scales about five times larger than the 2-point correlation. This result relies on a new estimator $\hr$ of pure phase-information in Fourier space, which can be interpreted as a parameter-free and scale-invariant tracer of filament-like structure. Based on simulated density fields we show how $\mdm$ can, in principle, be measured using $\hr$, given a suitably reconstructed density field.
\end{abstract}

\maketitle


\section{Introduction}\label{section_introduction}

The model of a flat and nearly scale-free universe dominated by dark energy and cold dark matter ($\Lambda$CDM), passed stringent empirical tests of the new millennium. The six free model parameters were found simultaneously consistent \citep{Komatsu2011} with the temperature fluctuations in the cosmic microwave background (CMB, \citealp{Larson2011}) measured by the Wilkinson Microwave Anisotropy Probe (WMAP, \citealp{Bennett2003a,Bennett2003b}), the baryon acoustic oscillations (BAOs) in the late-time large-scale structure (LSS) derived from galaxy redshift surveys (SDSS: \citealp{Percival2010}; 2dFGRS: \citealp{Percival2007}; WiggleZ: \citealp{Blake2011b}), and distance measurements based on type Ia supernovae (SNe, \citealp{Hicken2009,Kessler2009}).

This phenomenal success of the $\Lambda$CDM cosmology contrasts with our ignorance regarding the nature of its dark constituents. Crucial properties of these constituents, such as the particle mass of dark matter, are covertly imprinted in the sub-cluster structure of the LSS \citep{Smith2011,Schneider2012}. Yet, this information is not readily accessible to measurements. For one thing, the actual LSS is not directly observable due to the invisibility of dark matter, redshift-space distortions \citep{Kaiser1987}, and general relativistic effects \citep{Bonvin2011,Challinor2011,Yoo2009}. For another, the information in the LSS is masked by a random component originating from quantum state reduction in the primordial universe.

To filter out the random component, the observed LSS is usually subjected to statistical measures that are independent of cosmic randomness up to a volume-dependent shot noise, known as cosmic variance. In the case of a statistically homogeneous and isotropic universe, the infinite family of isotropic $n$-point correlation functions ($n$-PCFs) removes all randomness, but preserves all information \citep{Fry1985}. However, so far no \textit{finite} set of statistical measures is known, which exclusively and exhaustively describes the information imprinted in the cosmic density field. Most current studies bypass this issue by considering only the isotropic 2-PCF $\xi_2(r)$ or, equivalently, the power spectrum $p(k)$, where $r$ and $k$ denote the separation scale and wave-number. In doing so, important information is lost; e.g., subtle structural features, such as cosmic filaments, become indistinguishable from spherical features. Some studies improve on those drawbacks by invoking higher-order correlations \citep{Fry1978,Suto1994,Takada2003} and alternative statistical measures, such as the fractal correlation dimension \citep{Scrimgeour2012}, void distribution functions \citep{White1979}, and various shape-finders \citep{Babul1992,Luo1995,Sahni1998,Aragon-Calvo2007,Bond2010,Sousbie2011}. However, the benefit of these measures in addition to $\xi_2(r)$ is often limited, since they are heavily correlated to $\xi_2(r)$ in terms of ensembles. To truly avoid this issue one must refer to statistical estimators that only measure information not yet contained in $\xi_2(r)$ \citep[e.g.][]{Watts2003}.

The aim of this work is to introduce a new statistical estimator of the cosmic density field, which is based solely on the phases of the Fourier spectrum of the density field, but not on its amplitudes, since the latter are already fully captured via $\xi_2(r)$. This requirement combined with the requirement of statistical homogeneity and isotropy naturally leads to a measure, which we will call the line-correlation function $\hr$. We show that, in a limited sense, this function can be interpreted as a proxy for cosmic `filamentary' on length scales $2r$.

Unlike the 2-PCF, phase-correlations are independent of linear growth of LSS. Measures of phase-correlations such as $\hr$ are therefore particularly sensitive to the gravitational non-linear growth of the dark matter dominated cosmic web \citep{Watts2003}. In this work, we therefore chose to explore the dependence of $\hr$ on different `temperatures' of dark matter. Based on a series of large numerical $N$-body simulations, both with CDM and warm dark matter (WDM), we find that $\hr$ depends sensitively on the mass of dark matter particles $\mdm$. Our results suggest that $\hr$ constrains $\mdm$ an order of magnitude better than $\xi_2(r)$. Moreover, $\hr$ depends on $\mdm$ on scales about five times larger than $\xi_2(r)$
-- a pivotal result, since the complex baryon physics masking the footprint of dark matter properties becomes less important with increasing scales.

The article proceeds as follows. Section \ref{section_background} summarizes established concepts regarding cosmic structure and clarifies the meaning of $n$-PCFs and poly-spectra. Section \ref{section_linecorrelation} motivates and formally defines the line-correlation function and presents geometrical interpretations. A range of cosmological applications, namely the measurement of $\mdm$, is then considered in Section \ref{section_cosmology}, based on a series of $N$-body dark matter simulations. Section \ref{section_conclusion} summarizes the key results and discusses their potential application to observed data.

\section{Cosmic structure and correlation functions}\label{section_background}

This section reviews the statistical nature of cosmic LSS and summarizes the concepts of correlation functions and spectral analysis \citep[details in section 3 of][]{Bernardeau2002}. 

\subsection{Cosmic density field and its statistical symmetry}\label{subsection_density_field}

We consider a flat Euclidean universe, consistent with BAO measurements assuming a cosmological constant \citep{Percival2010,Blake2011b}, with a mass density field
\be\label{eq_def_density}
	\rho(\r) \equiv \frac{\d m(\r)}{\d V}\geq0,
\ee
where $m$ denotes the mass, $V$ the comoving volume, and $\r\in\R$ the position in $D$ spatial dimensions; for illustrative purposes we consider both $D=2$ and $D=3$. To simplify the notation, we omit the implicit time-dependance of $\rho$ in \eq{eq_def_density}. 

According to the Big Bang theory, $\rho(\r)$ evolved from a dense, maximally symmetric state under the action of physical laws that are spatially homogeneous and isotropic. Complex substructure then grew from seeds of reduced symmetry, known as quantum-fluctuations, caused by quantum state reduction in the inflating primordial universe \citep{Leon2011}.
In the current view, quantum state reduction decreases the spatial symmetry, but maintains homogeneity and isotropy in the sense that the outcome \textit{probabilities} of the process conserve the symmetry of the evolution operator \citep{McWeeny2002,Obreschkow2007}. This weaker, probabilistic symmetry is referred to as \textit{statistical} homogeneity and isotropy. The resulting conjecture that $\rho(\r)$ is statistically homogeneous and isotropic is known as the `cosmological principle' and is supported by modern redshift surveys (e.g.~SDSS: \citealp{Gong2010,Labini2010}; WiggleZ: \citealp{Scrimgeour2012}). We express the statistical homogeneity and isotropy explicitly by writing $\rho(\r)$ as
\be\label{eq_strong_cosmic_symmetry}
	\rho(\r) = \sum_{i}g_{i}(\t_i+R_i\r),
\ee
where $\t_i\in\mathds{R}^D$ are random translation vectors and $R_i\in O(D)$ are rotation matrices of the orthogonal group ($\det(R_i)=\pm1$). The generating functions $g_i(\r)\geq0$ are defined such that all cosmological information is encoded in $g_i(\r)$, while all quantum randomness is absorbed in the variables $\t_i$ and $R_i$. By definition, eq.~(\ref{eq_strong_cosmic_symmetry}) thus separates non-random variables $\{g_i(\r)\}$ from random ones $\{\t_i,R_i\}$. This separation is useful when constructing statistical measures that isolate the information.


Given the compelling observational evidence for the large-scale homogeneity, thus non-fractal structure, of the universe \citep[e.g.][]{Scrimgeour2012} we can define a universal average density $\bar{\rho}$ and the density perturbation field
\be\label{eq_def_density_perturbation}
	\dr \equiv \frac{\rho(\r)-\bar{\rho}}{\bar{\rho}}\geq-1.
\ee
This field then satisfies $\avg{\delta}_V\rightarrow0$ as $V\rightarrow\infty$, with $\avg{\delta}_V\equiv V^{-1}\int\d V \dr$ being the average density perturbation.

Let us comment on a few points. First, a corollary of the spatial homogeneity is that the total mass $\int_V\d V\rho(\r)=V\avg{\rho}_V$ is proportional to $V$ as $V\rightarrow\infty$. In other words, the fractal dimension converges to 3 in this limit \citep{Scrimgeour2012}. Second, it is a common misconception that statistical homogeneity and isotropy only concern large-scale ($\gtrsim\rm100 Mpc$) averages. For example, a universe with all mass concentrated around the nodes of a Cartesian grid with 1~Mpc spacings would satisfy $\avg{\rho}_V\rightarrow\bar{\rho}$ as $V\rightarrow\infty$, but violate statistical homogeneity as stated in eq.~(\ref{eq_strong_cosmic_symmetry}). Third, although our universe seems statistically homogeneous and isotropic, observational proxies of $\rho(\r)$, such as redshift-surveys, can violate this statistical symmetry. A famous example is the fingers-of-God effect \citep{Kaiser1987} originating from a Doppler-shift contamination in the observed redshifts -- a potentially useful feature for observational cosmology as emphasized by \cite{Raccanelli2012}.

\subsection{n-point correlation functions}\label{subsection_correlations}

Given a density field $\rho(\r)$ [eq.~(\ref{eq_strong_cosmic_symmetry})] that mixes information with random translations (homogeneity) and rotations (isotropy), how can we extract the information?

In a first step, statistical homogeneity is exploited by averaging over all translations. This is the key idea behind the correlation functions, which are spatial averages of product functions. The $n$-point density correlation function ($n$-PCF) is defined as
\be\label{eq_def_Xin}
	\Xi_n(\r_1,...,\r_{n-1}) \equiv \frac{1}{V}\int{\d^Dt\,\prod_{j=1}^{n}\delta(\t+\r_j)},
\ee
where $\r_n\equiv0$. In particular, the 2-PCF reads
\be\label{eq_def_Xi2}
	\Xi_2(\r) \equiv \frac{1}{V}\int{\d^Dt~\delta(\t)\delta(\t+\r)}.
\ee
In a second step, statistical isotropy is exploited by averaging over all rotations $R\in O(D)$. This leads to the `isotropic' $n$-PCFs,
\be\label{eq_def_xin}
	\xi_n(\mathcal{S}\{\r_1,...,\r_{n-1}\}) \equiv \avg{\Xi_n(R\r_1,...,R\r_{n-1})}_{R},
\ee
where $\mathcal{S}\{\r_1,...,\r_{n-1}\}$ denotes a unique representation of the shape defined by the $n$-point set $\{0,\r_1,...,\r_{n-1}\}$ regardless of its orientation. In the case of $n=2$, this shape reduces to the distance $r\equiv\Abs{\r_1}$. The resulting isotropic 2-PCF 
\be\label{eq_def_xi2}
	\xi_2(r) \equiv \avg{\Xi_2(R\r)}_{R}
\ee
is by far the most common statistical measure of LSS, as justified in Section \ref{subsection_gaussian_field}.

We emphasize that $\Xi_n$ and $\xi_n$ here refer to single realizations of the density field and not ensembles of fields, i.e.\ $\Xi_n\neq\ensavg{\Xi_n}$ and $\xi_n\neq\ensavg{\xi_n}$, where $\ensavg{~}$ denotes the ensemble average. Furthermore, these $n$-PCFs refer to the perturbation field $\dr$ rather than $\rho(\r)$. Our 2-PCF and 3-PCF are therefore identical to those called the `reduced' 2-PCF and 3-PCF by \cite{Peacock1999} and the `connected parts' of 2-PCF and 3-PCF by \cite{Bernardeau2002}. The family of the isotropic $n$-PCFs $\xi_n$ is statistically complete \citep{Fry1985} in that it contains all the information contained in the density field $\rho(\r)$. This information is contaminated by the cosmic variance $\ensavg{(\xi_n-\ensavg{\xi_n})^2}$, which can be calculated for any function $\xi_n$ \citep{Szapudi2001} and vanishes as $V\rightarrow\infty$.


\subsection{Fourier space representations}

Since the correlation functions $\Xi_n$ are convolution integrals over $\t\in\mathds{R}^D$, they can be computed more efficiently in Fourier space. Using the standard Fourier transform ${\rm FT}:\dr\mapsto\dk$ in cosmology and its inverse (IFT), and expressing all $\dr$ in \eq{eq_def_Xin} as ${\rm IFT}[\dk]$, we find (details in Appendix \ref{appendix_convolution})
\be\label{eq_connection}
\begin{split}
	& \Xi_n(\r_1,...,\r_{n-1}) = \left[\frac{V}{(2\pi)^D}\right]^{n-1}\!\!\int\d^Dk_1\,e^{i\k_1\cdot\r_1}\cdots \\
	& \qquad \times \int\d^Dk_{n-1}\,e^{i\k_{n-1}\cdot\r_{n-1}}~P_n(\k_1,...,\k_{n-1}),
\end{split}
\ee
where $\k\in\R$ is the wavevector and the complex-valued functions
\be
	P_n(\k_1,...,\k_{n-1})\equiv\hat{\delta}(\k_1)\cdots\hat{\delta}(\k_{n-1})\hat{\delta}(-\Sigma{\k_j})
\ee
are called `poly-spectra'. Thus, for any $n\geq2$, the correlation function $\Xi_n$ is equal to the IFT (generalized to $n-1$ variables) of the poly-spectrum $P_n$. The most common poly-spectra are the (real) power spectrum $P(\k)\equiv P_2(\k)$ and the bi-spectrum $B(\k,\q)\equiv P_3(\k,\q)$,
%
\bea
	P(\k) 	& ~=~ & \hat{\delta}(\k)\hat{\delta}(-\k) =  \big|\hat{\delta}(\k)\big|^2, \\
	B(\k,\q) 	& ~=~ & \hat{\delta}(\k)\hat{\delta}(\q)\hat{\delta}(-\k-\q). \label{eq_bqk}
\eea
Like in the case of the $n$-PCFs, the spectra $P(\k)\neq\ensavg{P(\k)}$ and $B(\k,\q)\neq\ensavg{B(\k,\q)}$ here refer to a single density field, not to ensembles thereof. These spectra are the FTs of the 2-PCF and 3-PCF, respectively,
\bea
	\Xi_2(\r) & = & \frac{V}{(2\pi)^D}\int\d^Dk~e^{i\k\cdot\r}~P(\k), \label{eq_2pt_convolution} \\
	\Xi_3(\r,\s) & = & \frac{V^2}{(2\pi)^{2D}}\!\iint\!\d^Dk\,\d^Dq\,e^{i(\k\cdot\r+\q\cdot\s)}\,B(\k,\q) \label{eq_3pt_convolution}.~~
\eea

The relations between $\Xi_n$ and $P_n$ imply similar relations between the isotropic correlation functions $\xi_n$ and rotationally symmetrized poly-spectra, called `isotropic' poly-spectra,
\be
	p_n(\mathcal{S}\{\k_1,...,\k_{n-1}\}) \equiv \avg{P_n(R\k_1,...,R\k_{n-1})}_{R}.
\ee

\begin{figure}
	\begin{center}
	\includegraphics[width=\columnwidth]{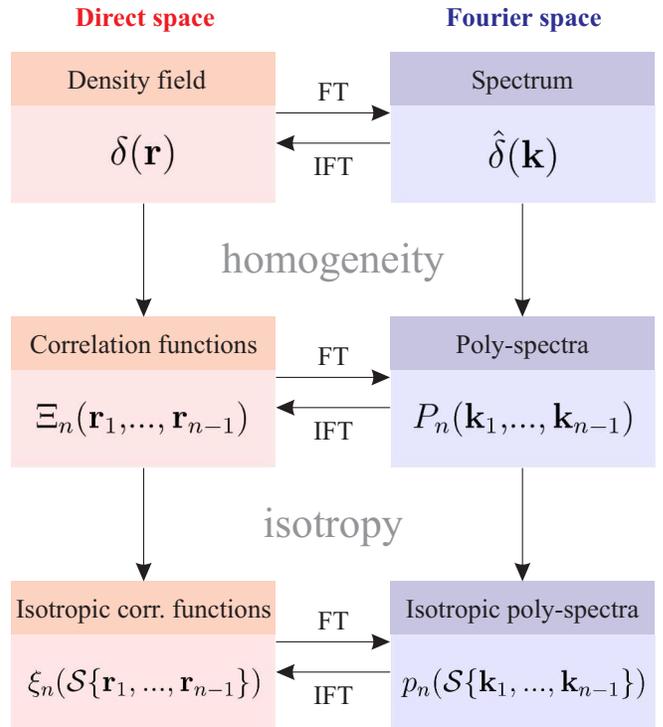}
	\caption{(Color online) Hierarchy of density field, correlations functions, and isotropic correlation functions together with their spectral equivalents. The irreversible mapping labeled `homogeneity' removes the random translations, which contain no information if $\dr$ is statistically homogeneous. Similarly, the irreversible mapping labeled `isotropy' removes the rotations, which contain no information if $\dr$ is statistically isotropic.}
	\label{fig_fourier_connection}
	\end{center}
\end{figure}

Fig.~\ref{fig_fourier_connection} depicts the hierarchy from the density field down to isotropic correlation functions and their equivalents in Fourier space. Because of their importance explicit expressions for $\xi_2$ and $\xi_3$ are given in Appendix \ref{appendix_explicit_expressions}.

\subsection{Cosmological importance of $\xi_2(r)$}\label{subsection_gaussian_field}

In the current view \citep{Bennett2011,Komatsu2011}, the primordial density fluctuations imprinted in the CMB are consistent with a Gaussian random field (GRF). A GRF results from a random (Poissonian) superposition of infinitely many plane or spherical waves with vanishing phase-correlation. The evidence for the Gaussianity of the CMB thus supports the physical interpretation that the primordial density fluctuations derive from de-correlated quantum fluctuations producing a bath of spherical sound waves. A key property of a GRF is that its information, i.e., its non-randomness, is entirely contained in the isotropic 2-PCF $\xi_2(r)$. Thus, as far as current measurements can tell, all the cosmological information of the CMB is contained in $\xi_2(r)$, or, equivalently, in its isotropic power spectrum $p(k)$.

\section{Phase-information and line-correlations}\label{section_linecorrelation}

This section introduces the `line-correlation' function, a new estimator of phase-information of cosmic structure.

\subsection{What is phase-information?}\label{subsection_phase_information}

For a statistically isotropic density field $\dr$, $\xi_2(r)$ in \eq{eq_def_xi2} contains the same information as the full 2-PCF $\Xi_2(\r)$. The reversible mapping between $\Xi_2(\r)$ in \eq{eq_2pt_convolution} and the amplitudes $\abs{\dk}$ then implies that \textit{$\xi_2(\r)$ measures all the cosmological information contained in the amplitude field $\abs{\dk}$}. All additional information, not captured by $\xi_2(r)$, must therefore reside in the phases-factors
\be
	\ek\equiv\frac{\dk}{\abs{\dk}}=e^{i\arg[{\dk}]}.
\ee
The information contained in these phase-factors is called phase-information and it can take the form of phase-phase correlations $\ensavg{\ek\hat{\epsilon}(\q)}$ and/or amplitude-phase correlations $\ensavg{\abs{\dk}\hat{\epsilon}(\q)}$. Unlike the primordial universe, the local universe does indeed contain a significant amount of phase-information, as evidenced by its clearly non-vanishing isotropic 3-PCF (e.g.~2dFGRS: \citealp{Croton2004,Gaztanaga2005}; SDSS: \citealp{Nichol2006,Marin2011}; numerical simulations: \citealp{Barriga2002}). Therefore, phase-information measurements of the late-time LSS promise to be a pivotal cosmological probe \citep{Watts2003}.

\subsection{Concept of line-correlations}\label{subsection_ideas}

\begin{figure*}
	\begin{center}
	\includegraphics[width=\textwidth]{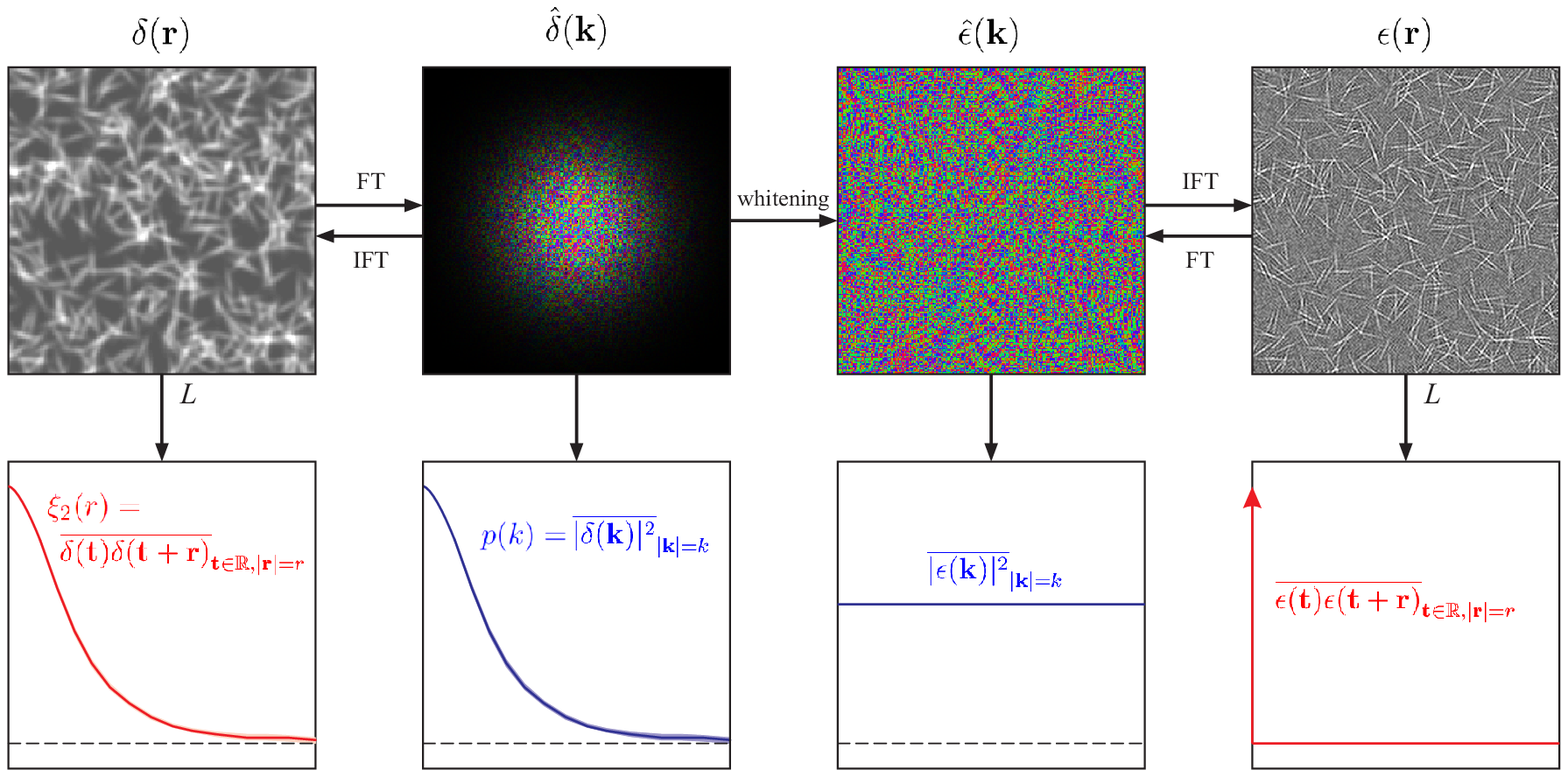}
	\caption{(Color online) Key idea of Section \ref{subsection_ideas}: to find an estimator of the density field $\dr$ that is uncorrelated to $\xi_2(r)$ up to residual correlations stemming from amplitude-phase correlations $\ensavg{\abs{\dk}\hat{\epsilon}(\q)}$, we remove all 2-point correlation from $\dr$. In Fourier space this is equivalent to suppressing all amplitude information, i.e., to the mapping $\dk\mapsto\ek\equiv\dk/|\dk|$. Any statistical measure depending only on $\ek$ or $\er={\rm IFT}(\ek)$ is then de-correlated from $\xi_2(r)$ for an ensemble of fields with vanishing amplitude-phase correlations. In this example, $\dr$ is a superposition of 200 randomly shifted and rotated 2D Gaussian distributions $\exp(-x^2/2\sigma_x^2-y^2/2\sigma_y^2)$ with $(\sigma_x,\sigma_y)=(0.006L,0.024L)$. The fields $\dr$, $\dk$, $\ek$, $\er$ are discretized using the scheme of Appendix \ref{appendix_discretization} with $N=300$ cells a side. The complex fields $\dk$ and $\ek$ are represented with brightness for amplitudes and hue-colors for phases.}
	\label{fig_method}
	\end{center}
\end{figure*}

A natural way to measure phase-information is to use higher-order correlations $\xi_n$ ($n\geq3$). However, this choice may be problematic since the estimators $\xi_n$ ($n\geq3$) and $\xi_2$ are strongly correlated in the sense that they correlate, i.e.~${\rm cov}(\xi_2,\xi_n)\neq0$, even if there are no amplitude-phase correlations, i.e.~$\ensavg{\abs{\dk}\hat{\epsilon}(\q)}=0$, across the considered ensemble of density fields. This strong correlation between $\xi_n$ ($n\geq3$) and $\xi_2$ is due to the fact that the poly-spectra $p_n$ depend directly on the amplitudes $\abs{\dk}$ in addition to the phases $\ek$. Thus, constraints on cosmological parameters obtained from $\xi_n$ ($n\geq3$) are generally strongly correlated to constraints obtained from $\xi_2$, which can lead to serious statistical difficulties. Alternative estimators, which are fully defined by phase-phase correlations $\ensavg{\ek\hat{\epsilon}(\q)}$ without explicit dependence on the amplitudes $\abs{\dk}$, are here called measures of \textit{pure} phase-information. Such estimators, must be defined exclusively upon the phase-factors $\ek$, or, equivalently, on the field $\er\equiv {\rm IFT}[\ek]$.

Fig.~\ref{fig_method} shows an example of a density field $\dr$ with the corresponding fields $\dk={\rm FT}[\dr]$, $\ek=\dk/\abs{\dk}$, and $\er={\rm IFT}[\ek]$. In this example, $\dr$ is a statistically homogeneous and isotropic 2D density field constructed on the basis of \eq{eq_strong_cosmic_symmetry}. The generating functions $g(\r)$ are identical elongated 2D Gaussian distributions. The mapping $\dk\mapsto\ek$ removes all information stored in the amplitudes $\abs{\dk}$ and therefore all 2-point correlations. Thus, the filamentary structure of $\er$ shown in Fig.~\ref{fig_method} exclusively represents phase-information of $\dr$.

By construction, $\er$ exhibits vanishing 2-point correlations for all $\r\neq0$ an thus the simplest meaningful measure of pure phase-information of $\dr$ must be based on 3-point correlations of $\er$. As illustrated in Fig.~\ref{fig_method}, the removal of 2-point correlation tends to collapse elongated structures to line segments. Therefore, the most natural 3-PCF to consider is that of three points on a straight line. For simplicity we chose these points to be equidistant. Using the explicit expressions for $\xi_3(r)\equiv\avg{\Xi_3(\r,-\r)}_{\abs{\r}=r}$ given in \eq{eq_xi3_first} and substituting $\dk$ for $\ek$, we then obtain the modified 3-PCF
\be
	\xi^\ast_3(r) = \frac{V^2}{(2\pi)^{2D}}\!\iint\!\d^{D\!}k\,\d^{D\!}q\,w_D(|\k-\q|r)\frac{B(\k,\q)}{\abs{B(\k,\q)}} \label{eq_ansatz2},
\ee
with the kernel
\be\label{eq_w}
	w_D(x) = \begin{cases}
		J_0(x), & \mbox{if } D=2, \\
		\sin(x)/x, & \mbox{if } D=3.
	\end{cases}
\ee
According to \eq{eq_connection}, $\xi^\ast_3(r)$ is identical to
\be
	\xi^\ast_3(r) = \avg{\epsilon(\t)\epsilon(\t+\r)\epsilon(\t-\r)}_{\t,|\r|=r}. \label{eq_ansatz3}
\ee

Unfortunately, $\xi^\ast_3(r)$ is an ill-defined function. (Mathematically, it is a distribution.) This can be seen when expressing $\epsilon$ in \eq{eq_ansatz3} as a discrete IFT. To do so, we adopt the standard discretization scheme explained in Appendix \ref{appendix_discretization} and illustrated in Fig.~\ref{fig_discretization}. In this scheme the density field is represented on a squared ($D=2$) or cubic ($D=3$) box with side-length $L$, $N^D$ grid cells spaced by $\Delta r=L/N$, and periodic boundary conditions. Its Fourier space becomes a regular box of side-length $2\pi/\Delta r$ and $N^D$ cells spaced by $\Delta k=2\pi/L$. The IFT then reads $\er=\sum_{\k}e^{i\k\cdot\r}~\ek$. A particular discretization is fully defined by $L$ and $\Delta r$, and any physical function $f(r)$ should converge both as $\Delta r\rightarrow0$ and $L\rightarrow\infty$. Yet, both requirements are violated by $\xi^\ast_3(r)$.

First, as $\Delta r\rightarrow0$, the number of independent modes with wavelengths shorter than a fixed $r$ diverges as $\Delta r^{-D}$, since $\max\{|\k|\}=\pi/\Delta r$. The $\sim\!\Delta r^{-D}$ modes with wavelengths shorter than the characteristic scale of $\dr$ exhibit vanishing amplitudes and thus random phase-factors $\ek$. Therefore, $|\er|=|\sum_{\k}e^{i\k\cdot\r}~\ek|\propto\Delta r^{-D/2}$, if $\Delta r\rightarrow0$ . According to \eq{eq_ansatz3}, $\xi^\ast_3(r)$ hence diverges as $\Delta r^{-3D/2}$ and becomes infinitely dominated by random noise. To avoid this divergence, we \textit{must} limit the number of high-frequency modes in \eq{eq_ansatz2}. By virtue of the Nyquist-Shannon theorem, it is natural to impose $|\k|\leq\pi/r$ and $|\q|\leq\pi/r$. In signal processing terminology, this mode-truncation is called a low-pass filter with a spherical top-hat kernel. With this modification $\xi^\ast_3(r)$ becomes independent of $\Delta r$, if $\Delta r<r/2$.

Second, as $L\rightarrow\infty$, we face a similar challenge: the density of modes per volume of Fourier space increases as $L^D$ due to the mode spacing $\Delta k=2\pi/L$. If $L$ is longer than the longest physical correlation lengths, then increasing $L$ corresponds to adding sub-modes with uncorrelated phases. Hence, we are again in the random phase case, where $|\er|=|\sum_{\k}e^{i\k\cdot\r}~\ek|\propto L^{D/2}$; $\xi^\ast_3(r)$ then diverges as $L^{3D/2}$. To avoid this, while keeping $\xi^\ast_3(r)$ dimensionless, we must divide $\xi^\ast_3(r)$ by $(L/r)^{3D/2}$.

\subsection{The isotropic line-correlation function $\hr$}

Based on the conceptual discussion of Section \ref{subsection_ideas}, we now define the `isotropic line-correlation function' of the density perturbation field $\dr$ as
\be\label{eq_def_linecorrelation}
	\hr \equiv \frac{V^{\frac{1}{2}}r^{\frac{3D}{2}}}{(2\pi)^{2D}}\!\!\!\!\!\!\!\underset{\substack{\Abs{\k},\Abs{\q}\leq\pi/r}}{\iint}\!\!\!\!\!\!\!\d^{D}\!k\,\d^{D}\!q\,w_D(|\k-\q|r)\frac{B(\k,\q)}{\Abs{B(\k,\q)}},
\ee
where $w_D(x)$ is given in \eq{eq_w}. Like the bispectrum $B(\k,\q)$, given in \eq{eq_bqk}, the line-correlation function $\hr\neq\ensavg{\hr}$ here refers to a particular realization of $\dr$ rather than an ensemble of fields. If some values of $\dk$ vanish, then $\Abs{B(\k,\q)}=0$ and $\hr$ diverges. In practice, this only happens for $\k=0$, where $\dk$ strictly vanishes by virtue of $\avg{\dr}=0$; for this case we adopt $\hat{\epsilon}(0)=\hat{\delta}(0)/\abs{\hat{\delta}(0)}\equiv0$. The discretized version of the function $\hr$ for the case of a periodic Cartesian grid with side-length $L$ reads (see Appendix \ref{appendix_discretization})
\be\label{eq_h_discretized}
	\hr = \left(\frac{r}{L}\right)^{\frac{3D}{2}}\!\!\sum_{\abs{\k},\abs{\q}\leq\pi/r}w_D(|\k-\q|r)\,\frac{B(\k,\q)}{\Abs{B(\k,\q)}}.
\ee

Given the definition of $\hr$ in eq.~(\ref{eq_def_linecorrelation}), a list of basic properties of can be derived:

\begin{enumerate}[(i)]
	\item $\hr$ measures statistically homogeneous and isotropic information in the field $\dr$.
	\item $\hr$ correlates to the 2-PCF and power spectrum only through amplitude-phase correlations, thus ${\rm cov}(\xi_2,\h)=0$ if $\ensavg{\abs{\dk}\hat{\epsilon}(\q)}=0$. In particular, for an ensemble of GRFs with random phases, $\xi_2$ and $\h$ are statistically independent -- a statement, which is not true for $\xi_2$ and $\xi_n$ ($n\geq3$).
	\item $\hr$ is invariant with respect to addition $\dr\mapsto\dr+c$ and multiplication $\dr\mapsto c\dr$ $(c\neq0)$, where $c$ is a real constant. By virtue of \eq{eq_def_density_perturbation}, $\hr$ of $\dr$ is therefore identical to $\hr$ of $\rho(\r)$.
	\item The mapping $\dr\mapsto\h(r)$ is non-linear with respect to the superposition of two density fields.
\end{enumerate}

\subsection{Physical interpretation of $\hr$}\label{subsection_physical_properties}

The correlation function $\hr$ is a measure of pure phase-information, i.e.~it depends only on phase-phase correlations $\ensavg{\ek\hat{\epsilon}(\q)}$. In turn, the 2-PCF $\xi_2(r)$ depends only on amplitude-amplitude correlations $\ensavg{\abs{\dk}\abs{\hat{\delta}(\q)}}$. The latter is often used as a measure of clustering. Similarly, $\hr$ can be interpreted as a measure of elongated structures, such as cosmic filaments, in a sense specified in the following.

\begin{figure*}
	\begin{center}
	\includegraphics[width=\textwidth]{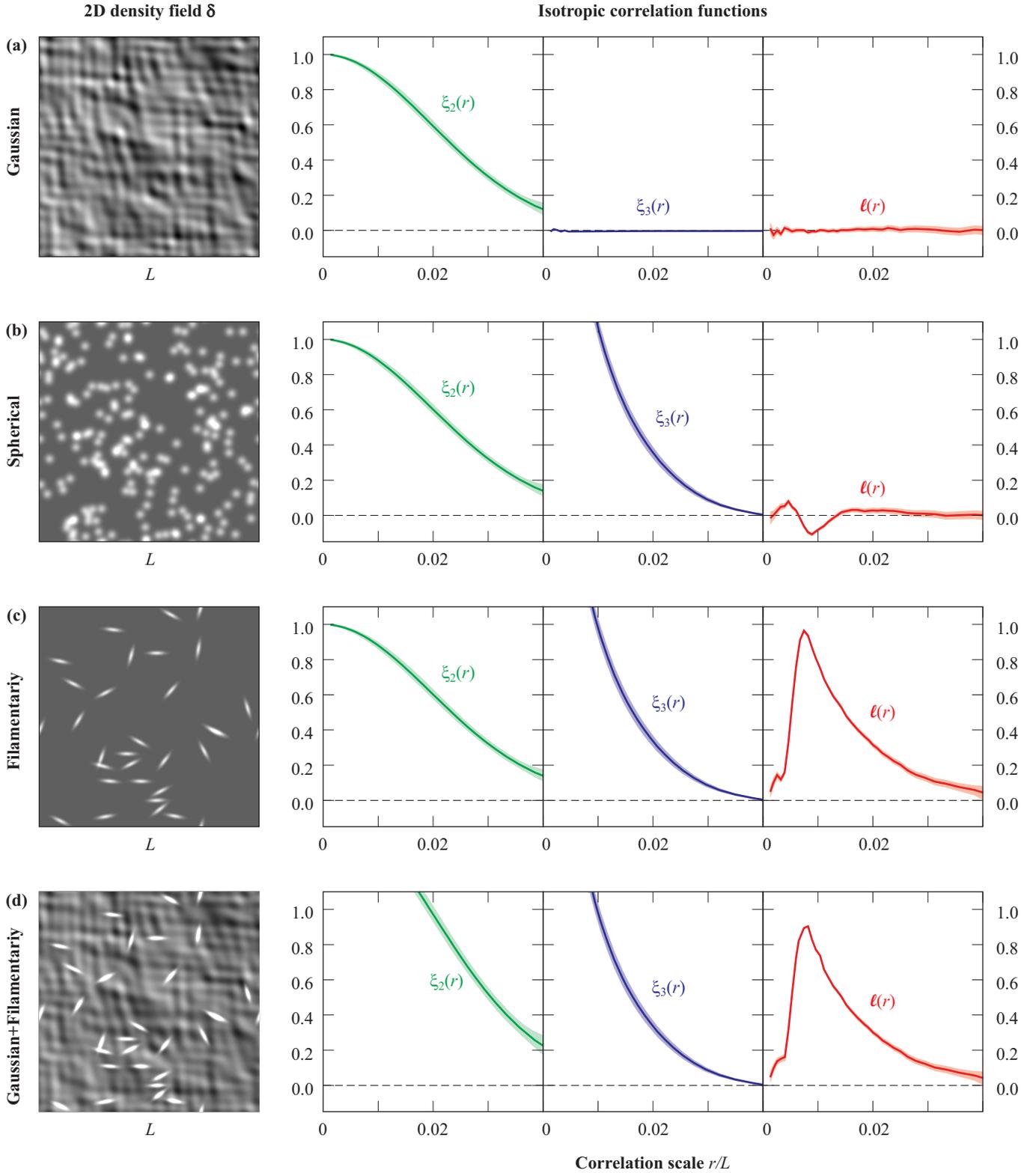}
	\caption{(Color online) Example of three 2D density fields $\dr$ with the corresponding correlation functions $\xi_2(r)$, $\xi_3(r)$, and $\hr$. The shaded envelopes of the solid lines represent 67\% confidence intervals. The density fields are chosen such that they all yield a similar $\xi_2(r)$ and such that two of them yield a similar $\xi_3(r)$. (a) GRF constructed by superposing 350 randomly oriented plane waves; (b) a superposition of 200 symmetric 2D-Gaussian distributions $\exp(-r^2/2\sigma^2)$ with $\sigma=0.014L$; (c) a superposition of 30 randomly oriented asymmetric 2D Gaussian distributions $\exp(-x^2/2\sigma_x^2-y^2/2\sigma_y^2)$ with $(\sigma_x,\sigma_y)=(0.006L,0.024L)$; (d) density field is equal to the sum of the density fields in examples (a) and (c).}
	\label{fig_examples}
	\end{center}
\end{figure*}

Fig.~\ref{fig_examples} shows four examples of a 2D random field $\dr$, constructed by using \eq{eq_strong_cosmic_symmetry}: (a) a GRF made of plane waves, (b) a random field with circular kernels, (c) a random field using filament-like kernels, and (d) a superposition of the GRF (a) and the filamentary field (c). The fields (a)--(c) are designed to exhibit the same 2-PCF $\xi_2(r)$, thus illustrating that $\xi_2(r)$ cannot distinguish between wave-like, spherical, and filamentary substructure (see also Fig.~1 by \citealp{Coles2005}). In contrast, $\xi_3(r)$ vanishes for the GRF, but still exhibits a similar shape for the spherical and filamentary fields. In principle, the full isotropic 3-PCF $\xi_3(\abs{r},\abs{s},\angle{\r,\s})$ can distinguish spherical from filamentary structure \citep[e.g.][]{Nichol2006}, but its geometrical interpretation is difficult. Only the line-correlation function $\hr$ clearly separates the filamentary density field (c) from the fields (a) and (b). Moreover, the shape of $\hr$ describes the straight filaments quantatively: $\hr$ exhibits a bump, $\h(r)>0.2$, roughly on the interval $r\in[\sigma_x,\sigma_y]$ between the short and the long characteristic filament radius (see caption of Fig.~\ref{fig_examples}) . Hence, for a density field composed exclusively of straight filaments, $\hr$ measures the filamentarity on length scales $2r$. This feature of $\hr$ remains true, even if the filamentary field (c) is superposed with the GRF (a), such as illustrated in Fig.~\ref{fig_examples}d.

\begin{figure*}
	\includegraphics[width=\textwidth]{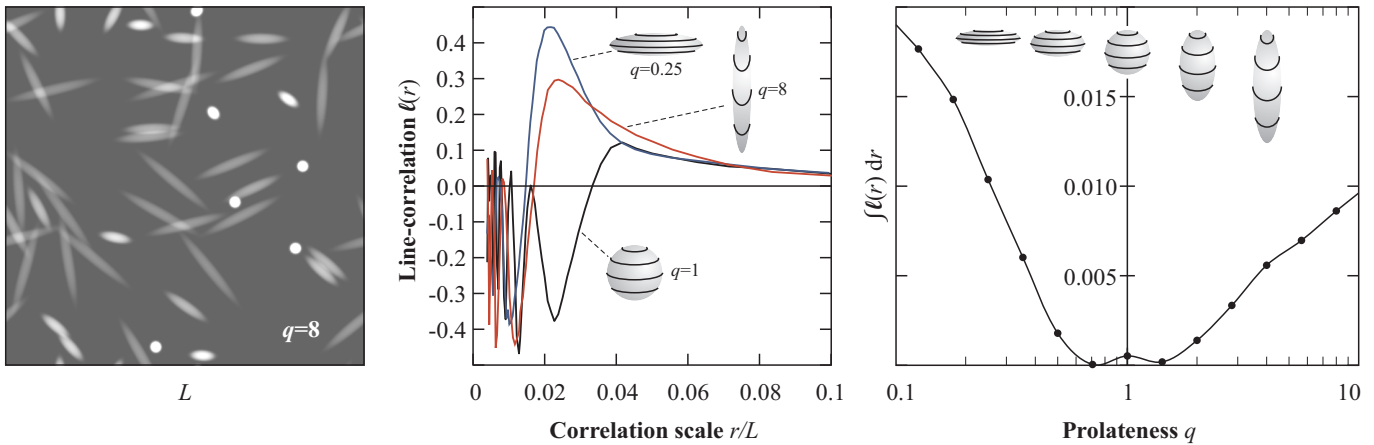}
	\caption{(Color online) Illustration of how the function $\hr$ captures aspherical structure. LEFT: projection of a 3D density field $\dr$ consisting of 50 randomly translated and rotated identical spheroids. These spheroids are characterized by the ``prolateness'' $q$, defined as the ratio between the pole-radius $r_p$ and the equator-radius $r_e$, and the average radius of $(r_pr_e^2)^{1/3}=0.03L$. MIDDLE: functions $\hr$ corresponding to the density fields with q=0.25 (oblate spheroids), q=1 (spheres), and q=8 (prolate spheroids). In the oblate and prolate case, $\hr$ exhibits a maximum between $r_p$ and $r_e$. RIGHT: Integral of $\hr$ as a function of $q$.}
	\label{fig_varying_prolateness}
\end{figure*}

A more systematic account of how aspherical substructure is imprinted in $\hr$ is provided in Fig.~\ref{fig_varying_prolateness}. This analysis relies on a 3D density field constructed via \eq{eq_strong_cosmic_symmetry}, with the generating functions $g(\r)$ being 50 identical spheroids. These spheroids are fully characterized by their prolateness $q$, defined as the ratio between the pole-radius $r_p$ and the equator-radius $r_e$, and by their average radius $r_0=(r_pr_e^2)^{1/3}$. The left panel in Fig.~\ref{fig_varying_prolateness} shows a projection of the 3D density field for the case of $q=8$. The middle panel shows $\hr$ for three selected values of $q$, confirming that $\hr$ is positive for $r>\min\{r_e,r_p\}$ and peaks in the interval $r\in[r_e,r_p]$. For $r<\min\{r_e,r_p\}$, $\hr$ undergoes a series of oscillations. In the case of perfectly spherical ($q=1$) substructure $\hr$ does not vanish; however, the integral $\int\h(r)\d r$ nearly vanishes for $q=1$ and increases for both $q<1$ and $q>1$. We thus see that $\hr$ is particularly sensitive to aspherical substructure on scales $\sim2r$, be it oblate, such as cosmic sheets, or prolate, such as the more common cosmic filaments.

\begin{figure}[b]
	\includegraphics[width=\columnwidth]{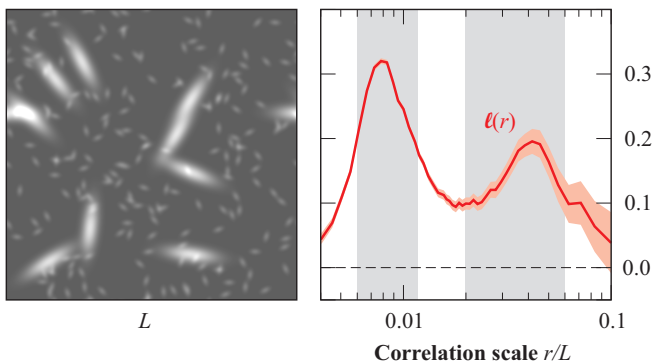}
	\caption{(Color online) Illustration of $\hr$ measuring two sizes of filaments simultaneously. LEFT: 2D density field $\dr$ composed of randomly shifted and rotated distributions $\exp(-x^2/2\sigma_x^2-y^2/2\sigma_y^2)$, where 50 distributions use $(\sigma_x,\sigma_y)=(0.006L,0.012L)$ and 10 use $(\sigma_x,\sigma_y)=(0.02L,0.06L)$. RIGHT: The corresponding function $\hr$ displays two peaks approximately centered within the two invervals $[\sigma_x,\sigma_y]$ (gray shadings). 1-sigma shot noise uncertainties of $\hr$ are represented by the light-red envelope.}
	\label{fig_double_filament}
\end{figure}

The line-correlation function $\hr$ can also be used to characterize a mixture of aspherical structures. An illustration for the case of a 2D density field is shown in Fig.~\ref{fig_double_filament}. This density field is again constructed via \eq{eq_strong_cosmic_symmetry}, by randomly superposing two sizes of filaments. In this case $\hr$ is double-peaked with each peak approximately measuring the width and length of one type of filament.

\begin{figure}
	\includegraphics[width=\columnwidth]{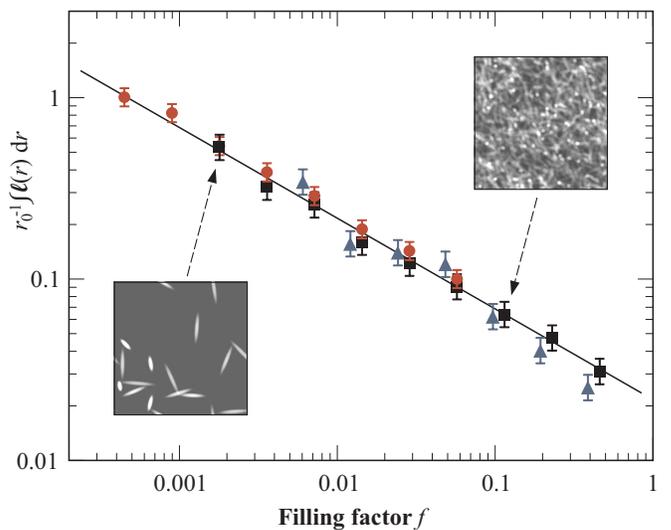}
	\caption{(Color online) Illustration of \eq{eq_integral_scaling} for the case of a 3D density field composed of spheroidal functions with $q=8$ (as in Fig.~\ref{fig_varying_prolateness}, left). The points are computational results using $r_0=0.015L$ (dots), $r_0=0.030L$ (squares), and $r_0=0.045L$ (triangles). Error bars represent 1-sigma shot noise uncertainties for the chosen numerical discretization. The solid line is the power-law of \eq{eq_integral_scaling} with $\kappa$ fitted to the data.}
	\label{fig_varying_filling_factor}
\end{figure}

Next, we investigate the dependence of $\hr$ and $\int\h(r)\d r$ on the size and density of substructure, e.g.~the length and density of filaments. Let us consider a density field constructed via \eq{eq_strong_cosmic_symmetry} by superposing $m$ identical generating functions $g(\r)$ of a characteristic linear scale $r_0$ and a characteristic volume $V_0$, e.g., spheroids with $r_0=(r_pr_e^2)^{1/3}$ and $V_0=(4\pi/3)r_0^3$. We define the filling-factor of this density field as $f\equiv mV_0/L^3$. If the filling factor is increased by increasing the number of objects $m$, then the increasing number of random translations and rotations in \eq{eq_strong_cosmic_symmetry} amplifies the phase-noise as $\sqrt{m}$. Hence the amplitude of the line-correlation $\hr$ falls (on average) as $1/\sqrt{m}$. If, on the other hand, the characteristic scale $r_0$ is changed by a factor $u$ while maintaining $f$ constant, i.e., by varying the number of objects per unit volume, the situation looks as follows. Let $\tilde{r}_0\equiv u r_0$ be the new length scale, and $\tilde{\delta}(\r)$ the corresponding density field. Since $\hr$ is independent of the box size $L$ as $L\rightarrow\infty$, we may choose the new density field $\tilde{\delta}(\r)$ to be defined on a cubic volume with side-length $\tilde{L}=u L$, such that the number of objects $\tilde{m}$ in the total volume remains the same, i.e., $\tilde{m}=m$. In this way we find $\tilde{\delta}(u\r)=\delta(\r)$, given the same choice of random translations and rotations for $\dr$ and $\tilde{\delta}(\r)$. The line-correlation associated with $\tilde{\delta}$ then becomes $\tilde{\h}(r)=\h(r/u)$. As a consequence, if in a density field $\dr$ with line-correlation $\hr$ the substructure-scale $r_0$ is stretched by a factor $u$ and the filling factor $f$ is varied by a factor $v$, the line-correlation scales as
\be\label{eq_h_scaling}
	\tilde{\h}(r) = v^{-1/2}~\h\left(\frac{r}{u}\right)
\ee
It follows that the rescaled function $\sqrt{f}\,\h(r/r_0)$ is independent of $f$ and $r_0$ (up to shot noise) and thus characteristic of a particular type of substructure, e.g., spheroids of a certain prolateness $q$. This also implies that
\be\label{eq_integral_scaling}
	r_0^{-1} \int\h(r)\,\d r = \kappa f^{-1/2}
\ee
with $\kappa$ being a constant depending on the shape of the substructure, but not on its scale and filling factor. This relation is strongly supported by the numerical example in Fig.~\ref{fig_varying_filling_factor}.

In summary, $\hr$ is sensitive to aspherical substructure on scales $2r$. In the special case of a filamentary field, $\hr$ measures the characteristic scales of the filaments. The dependance of $\hr$ on the prolateness of aspherical substructure is shown in Fig.~\ref{fig_varying_prolateness}. The dependance of $\hr$ on the size and filling factor of substructure is given in \eq{eq_h_scaling}, and illustrated in Fig.~\ref{fig_varying_filling_factor}.

\section{Application to cosmic structure}\label{section_cosmology}

This section studies the line-correlation $\hr$ of simulated cosmic density fields, using both CDM and WDM, i.e., dark matter with a finite particle mass.

\subsection{Cosmological simulation}\label{subsection_simulation}

We have run a suite of cosmological $N$-body simulations, following the formation and evolution of LSS in a cubic box of comoving side-length $L=100\hmpc$ containing $512^3\approx1.34\cdot10^8$ particles, from an initial redshift of $z_{\rm init}=199$ to $z=0$. Following \citet{Komatsu2011}, we adopt matter and dark energy density parameters of $\Omega_0=0.273$ and $\Omega_{\Lambda}=0.727$, a Hubble parameter of $h=0.705$ (defined by $H_0=100\,h\rm~km\,s^{-1}\,Mpc^{-1}$), a primordial spectral index of $n_{\rm spec}=0.95$ and a normalization $\sigma_8=0.812$. 

Initial conditions were created using standard techniques \citep[e.g.][]{Power2003} -- a statistical realization of a GRF is generated in Fourier space, with variance given by the linear matter power spectrum, and the Zel'dovich approximation is used to compute initial particle positions and velocities. The power spectrum for the CDM model is obtained by convolving the primordial isotropic power spectrum $p(k) \propto k^{n_{\rm spec}}$ with the transfer function appropriate for our chosen set of cosmological parameters, computed using the Boltzmann code {\texttt{CAMB}} \citep[cf.][]{Lewis2000}. Following \citet{Bode2001}, the initial power spectra for our WDM models are obtained by filtering the CDM power spectrum with the transfer function
\be\label{eq_transfer_function}
  T_{\rm WDM}(k) = \left(\frac{p_{\rm WDM}(k)}{p_{\rm CDM}(k)}\right)^{1/2}=\left[1+(\mu\,k)^{2\nu}\right]^{-5/\nu},
\ee
where $\mu$ is a function of the WDM particle mass \citep[labeled `$\alpha$' in eq.~(A9) of][]{Bode2001} and $\nu=1.2$ is a numerical constant. Eq.~(\ref{eq_transfer_function}) mimics the free-streaming of WDM particles by preferentially suppressing high-frequency modes.

All simulations were run using the parallel TreePM code {\small GADGET2} \citep{Springel2005} with constant comoving gravitational softening $\epsilon=4\rm~kpc$ and individual and adaptive time steps for each particle, $\Delta t = \eta \sqrt{\epsilon/a}$, where $a$ is the magnitude of a particle's gravitational acceleration and $\eta=0.05$ determines the accuracy of the time integration.

In order to study spatial correlations, the particles are discretized onto a regular grid of $N^3$ cells (here $N=400$), as described in Appendix \ref{appendix_discretization}. The correlation functions are then computed via eqs.~(\ref{eq_correlations_discretized}).

\subsection{Line-correlation $\hr$ in a $\Lambda$CDM universe}\label{subsection_cdm}

\begin{figure}
	\includegraphics[width=\columnwidth]{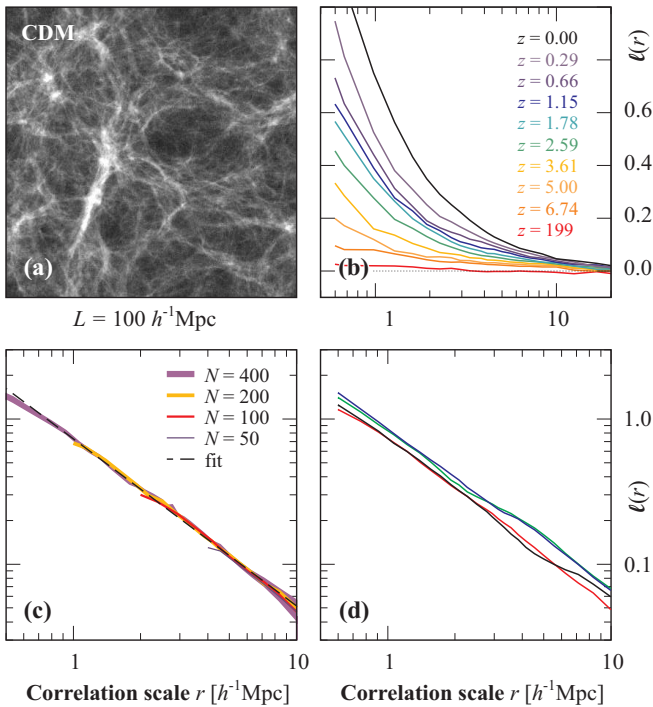}
	\caption{(Color online) Illustration of the function $\hr$ in the case of a virtual cosmic density field, simulated as described Section \ref{subsection_simulation}. (a) Plane-projection of the 3D field $\dr$ at a simulation time of $13.7\rm~Gyr$, i.e., at $z=0$. (b) Cosmic evolution of $\hr$, starting from a Gaussian density field at $z=z_{\rm init}=199$, where $\h(r)\approx0$. (c) Solid lines show the measured $\hr$ for four different discretizations of the simulation volume into $N^3$ cells with $N=50,100,200,400$; the dashed line represents the power-law fit of \eq{eq_powerlaw}. (d) $\hr$ for three different random realizations of the simulation to illustrate the effect of cosmic variance for a volume of $(100\,\hmpc)^3$.}
	\label{fig_example_cdm}
\end{figure}

Fig.~\ref{fig_example_cdm}a shows a projection of a simulated CDM field at a cosmic time of $13.7\rm~Gyr$, i.e., a redshift $z=0$, while Fig.~\ref{fig_example_cdm}b displays the cosmic evolution of $\hr$ starting at a cosmic time corresponding to $z_{\rm init}=199$. At this early time, where the universe still closely resembles the initial GRF, $\hr$ nearly vanishes. However, as the universe evolves, $\hr$ monotonously grows for all values of $r$, hence uncovering a continuous and monotonous growth of phase-correlations on all scales accessible to the simulation ($r\leq L/4=25\hmpc$). We interpret this rise of $\hr$ as a growing presence of gravity-induced aspherical structure, namely tidal cosmic filaments.

We find that the late-time ($z=0$) line-correlation $\hr$ is well approximated by the power-law
\be\label{eq_powerlaw}
	\h_{\rm CDM}(r) = 0.73\,\left(\frac{r}{\hmpc}\right)^{-1.15},
\ee
shown as dashed-line in Fig.~\ref{fig_example_cdm}c. A power-law behavior had to be expected for it is equivalent to $\hr$ being scale-free in the sense that $\h(r_1)/\h(r_2)$ is constant if $r_1/r_2$ is constant -- a general feature of LSS on scales $10\rm~kpc<r<10\rm~Mpc$ (e.g.~the weak-lensing analysis of the Sloan Digital Sky Survey (SDSS), see Fig.~10~of \citealp{Sheldon2004}). Interestingly, the power-law of \eq{eq_powerlaw} is less steep than the observed cosmic 2-PCF $\xi_2(r)\propto r^{-1.79}$ \citep{Sheldon2004}, consistent with the interpretation that $\hr$ senses structural features extending to comparatively large scales, such as filaments.

Fig.~\ref{fig_example_cdm}c also illustrates the numerical convergence of $\hr$ for an increasing number of grid cells $N^3$ (different solid lines). The convergence is ensured by the normalization factor in front of the integral in \eq{eq_def_linecorrelation} as explained in Section \ref{subsection_ideas}.

Finally, Fig.~\ref{fig_example_cdm}d explores the effect of cosmic variance on $\hr$. The four solid lines correspond to four different random realizations of the initial density field. We notice that cosmic variance dominantly affects the amplitude of the power-law, but only marginally affects its slope. From these examples we estimate that, for a box size $L=100\hmpc$, the normalization of $0.73$ in \eq{eq_powerlaw} has a statistical uncertainty of about 10\%, while the exponent of $-1.15$ is accurate to about $3\%$.

\subsection{Measuring the `temperature' of dark matter}\label{subsection_measuring_temperature}

\begin{figure}
	\includegraphics[width=\columnwidth]{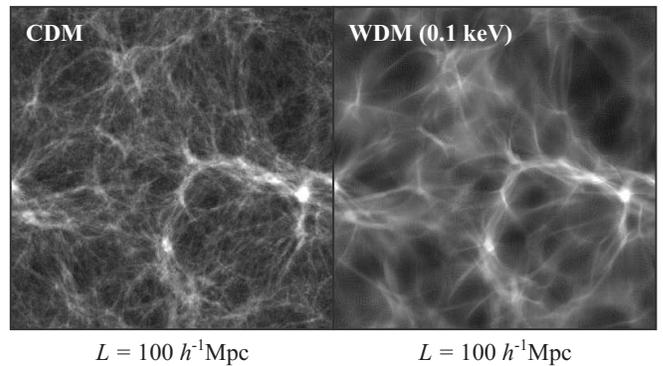}
	\caption{Plane-projection of two simulated 3D fields $\dr$ at a simulation time of $13.7\rm~Gyr$, i.e., at $z=0$. The two fields rely on identical primordial initial conditions, using CDM and WDM at $\mdm=0.1\rm~kev$, respectively. The WDM field seems smoother because of the suppression of short modes via \eq{eq_transfer_function}. The continuous transition between the two panels reflects the periodic boundary conditions of the two boxes.}
	\label{fig_cdm_and_wdm}
\end{figure}

What cosmological information can be extracted from a local measurement of $\hr$? Since $\hr$ only depends on correlations between the phases $\ek$, it is strictly insensitive to linear growth, defined as a uniform growth of the amplitudes $\abs{\dk}$. Therefore the physics associated with linear growth, such as that dictating the baryon acoustic scale, remains invisible to $\hr$. However, $\hr$ depends on the physics associated with non-linear growth, namely on local gravitational interactions and hence on the properties of dark matter. We therefore chose to examine the variations of $\h(r)$ with the dark matter particle mass $\mdm$, using the WDM simulations described in Section \ref{subsection_simulation}. An illustration of two analogous density fields with CDM and WDM is provided in Fig.~\ref{fig_cdm_and_wdm}.

\begin{figure}
	\includegraphics[width=\columnwidth]{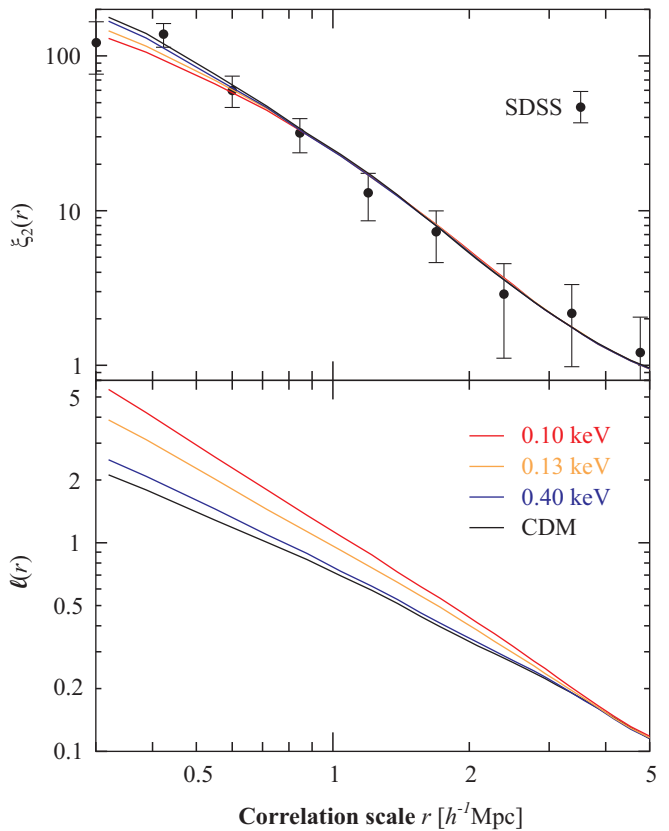}
	\caption{(Color online) 2-PCF $\xi_2(r)$ and line-correlation $\hr$ associated with four different dark matter particle masses. These examples rely on simulations with identical initial conditions. The data points are the galaxy-mass correlation function measured from weak-lensing in SDSS \citep{Sheldon2004}.}
	\label{fig_varying_temperature}
\end{figure}


Fig.~\ref{fig_varying_temperature} shows the functions $\xi_2(r)$ and $\hr$ for four values of $\mdm$, here expressed in units of energy, e.g.~$1\rm~keV\cong1{\rm~keV}\,c^{-2}\approx1.78\cdot10^{-33}\rm~kg$. Fig.~\ref{fig_varying_temperature} confirms that $\xi_2(r)$ only weakly depends on our array of $\mdm$, and that this dependance is restricted to scales $r<1\,\hmpc$, in good agreement with several recent studies \citep[e.g.][]{Smith2011}. By contrast, $\h(r)$ exhibits a much stronger dependance on $\mdm$, and this dependence extends to scales of about $5\,\hmpc$.

The $\mdm$-dependencies displayed in Fig.~\ref{fig_varying_temperature} can be reversed to infer $\mdm$ from $\xi_2(r)$ and $\h(r)$. To simplify the notations, let $f(r)$ be a generic placeholder for $\xi_2(r)$ and $\h(r)$. When analyzing the variations of $\log_{10}f(r)$ as a function of $\mdm$ (for $r>0.5~\hmpc$ and $\mdm<1\rm~keV$), we find them to be linear in $\mdm^{-1}$ within the uncertainties of cosmic variance. It follows that, at any $r$, the true particle mass $\mdm^{\rm true}$ can be detected against a hypothetical particle mass $\mdm^{\rm hyp}$, e.g.~against CDM ($\mdm^{\rm hyp}=\infty$), with a signal-to-noise ratio proportional to $\Delta\mdm^{-1}=(\mdm^{\rm hyp})^{-1}-(\mdm^{\rm true})^{-1}$. In other words, this signal-to-noise ratio can be expressed as $s(r)\abs{\Delta\mdm^{-1}}$, where $s(r)$ denotes the signal-to-noise per unit of $\mdm^{-1}$. We here approximate $s(r)$ as
\be\label{eq_sr}
	s(r) = \frac{\abs{\log_{10}f_{\rm WDM\,0.1keV}(r)-\log_{10}f_{\rm CDM}(r)}}{10~{\rm keV}^{-1}~\mathcal{N}(r)},
\ee
where $\mathcal{N}(r)$ represents the noise, defined as the standard deviation of $\log_{10}f(r)$ due to cosmic variance. We estimate $\mathcal{N}(r)$ as the root-mean-square of $\log_{10}f_{\rm CDM}(r)$ over four random realizations of a CDM run. This numerical estimate is more reliable than analytical estimates based on the number of Fourier modes, if the latter are phase-correlated. The normalization factor in the denominator of \eq{eq_sr} comes from the fact that the values of $\mdm^{-1}$ in CDM and WDM at $0.1\rm~keV$ differ by $10~{\rm keV}^{-1}$.

When measuring $\mdm$ based on $f(r)$, we consider $\Delta\mdm^{-1}(r)$ to be the difference between a measurement of $\mdm^{-1}$, at a specific scale $r$, and its true value. In the Gaussian approximation, the probability distribution $\phi(r)$ of $\Delta\mdm^{-1}(r)$ is then proportional to $\exp{[-\Delta\mdm^{-2}s^{2}(r)/2]}$. Note that the exponent is dimensionless, as it should be. Combining the measurements on the scales $r\in[r_{\rm min},r_{\rm max}]$ associated with independent Fourier modes $k=\pi/r$, the total probability distribution of $\Delta\mdm^{-1}$ becomes $\phi(r_{\rm min})\cdot...\cdot\phi(r_{\rm max})$. Hence the standard-deviation of a measurement of $\mdm^{-1}$ on scales $r\in[r_{\rm min},r_{\rm max}]$ becomes
\be\label{eq_sigma}
	\sigma(r_{\rm min},r_{\rm max}) = \left[\sum_{r=r_{\rm min}}^{r_{\rm max}} s^2(r)\right]^{-\frac{1}{2}}.
\ee

\begin{figure}
	\includegraphics[width=\columnwidth]{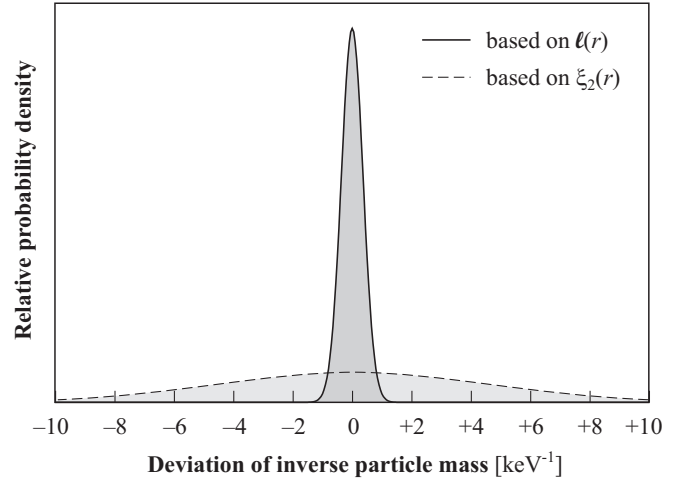}
	\caption{Probability distribution of the difference between the measured inverse particle mass $\mdm^{-1}$ and its true value. The solid line represents a phase-correlation measurement via $\hr$, while the dashed line represents an amplitude-correlation measurement via $\xi_2(r)$. Both rely on the scales $0.5\,\hmpc\leq r\leq5\,\hmpc$ in a cubic volume $V=(100\,\hmpc)^3$.}
	\label{fig_mass_probability}
\end{figure}

Numerical estimates of $\sigma(r_{\rm min},r_{\rm max})$ for various intervals $[r_{\rm min},r_{\rm max}]$ and for both correlation functions ($\xi_2$ and $\h$) are given in Tab.~\ref{tab_sigma100}. Fig.~\ref{fig_mass_probability} shows the probability distribution of $\Delta\mdm^{-1}$ for the case of a measurement based on the interval $[0.5~\hmpc,5~\hmpc]$.

The following points are worth stressing. First, the values in Tab.~\ref{tab_sigma100} and probability distributions in Fig.~\ref{fig_mass_probability} correspond to a perfect measurement of the density field $\dr$, since they only account for the fundamental limitations associated with cosmic variance. They ignore potentially large measurement uncertainties and observational biases. Second, the values in Tab.~\ref{tab_sigma100} are specific to our box-size $L=100~\hmpc$. The noise $\mathcal{N}(r)$ in \eq{eq_sr} scales as $V^{-1/2}$, where $V=L^3$ is the volume of the considered density field. Hence standard-deviations of $\mdm^{-1}$ for any other volume $V$ can be obtained as $[(100~\hmpc)^3/V]^{1/2}\sigma$. Third, the standard-deviation of $\mdm$ rather than $\mdm^{-1}$ is given by $\mdm^2\sigma$. 

\begin{table}[b]
	\centering
	\normalsize
	\begin{tabular}{ccc}
	\hline\hline \\ [-1.5ex]
	$[r_{\rm min},r_{\rm max}]/\!\hmpc$ & $\sigma$ for $\xi_2(r)$ & $\sigma$ for $\hr$ \\ [1.0ex]
	\hline \\ [-1.5ex]
	$[0.50,2.00]$	&	3.3		&	0.3	\\
	$[1.00,4.00]$	&	$\gg10$	&	0.6	\\
	$[0.50,1.00]$	&	2.3		&	0.3	\\
	$[1.00,2.00]$	&	$\gg10$	&	0.5	\\
	$[2.00,4.00]$	&	$\gg10$	&	1.7	\\ [1.5ex]
	\hline
	\end{tabular}
	\caption{\upshape\raggedright Numerical values of the standard deviations $\sigma(r_{\rm min},r_{\rm max})$ of a measurement of $\mdm^{-1}$ in a cosmic volume $V=(100~\hmpc)^3$.}
	\label{tab_sigma100}
\end{table}

The key result from this analysis is that, in the late-time universe, the particle mass of dark matter $\mdm$ is much better constrained by phase-phase correlations, measured via $\hr$, than by amplitude-amplitude correlations, measured via $\xi_2(r)$ or $p(k)$. Furthermore, $\mdm$ affects scales five times larger in $\hr$ than in $\xi_2(r)$. This finding is crucial, since the smaller the scale required to measure $\mdm$, the more the result will be entangled with the uncertainties of complex baryon physics, such as feedback from supernovae, active black holes, and photo-ionization \citep{Kim2012}.

The reason for this advantage of $\hr$ over $\xi_2(r)$ is twofold. First, phase-phase correlations vanish in the primordial CDM/WDM power spectrum and are strictly independent of linear growth. They therefore represent exquisite tracers of non-linearly grown structure, which in turn depends substantially on $\mdm$. This supports the finding that $\hr$ is more sensitive on $\mdm$ than $\xi_2(r)$ for a given scale $r$. Second, $\hr$ is particularly sensitive to cosmic filaments, while $\xi_2(r)$ traces spherical structures, such as clusters. However, cosmic filaments extend to larger scales than galaxy clusters and they seem to better preserve the primordial free-streaming scale (see Fig.~\ref{fig_cdm_and_wdm}), perhaps because filaments are second order effects of the gravitational field \citep{Pen2012}. This might explain why $\hr$ traces $\mdm$ to larger scales than $\xi_2(r)$.

Caution is nonetheless indicated. Despite the advantage of $\hr$ over $\xi_2(r)$, the case presented in Fig.~\ref{fig_varying_temperature} suggests that particle masses $\mdm>1\rm~keV$ will remain difficult to distinguish from CDM on scales larger than $0.5~\hmpc$. However, particles lighter than $1-2\rm~keV$ seem inconsistent with the Lyman-$\alpha$ forest \citep[][]{Boyarsky2009}. To probe particle masses $\gg1\rm~keV$, $\hr$ will have to be measured on smaller scales \citep[eq.~(3) in][]{Schneider2012}, not yet studied in this work. Those scales will still be larger than those required by $\xi_2(r)$, but baryon physics will become important.

\section{Conclusion}\label{section_conclusion}

\subsection{Three key messages}

\subsubsection{$\hr$ -- a measure of pure phase-information}

This work introduced the isotropic line-correlation function $\hr$, defined for a density field $\dr$ via \eq{eq_def_linecorrelation}. Unlike conventional $n$-PCFs, $\hr$ is defined exclusively upon the spectral phases $\ek$. Thus $\hr$ only measures phase-phase correlations $\ensavg{\ek\hat{\epsilon}(\q)}$. By contrast, the 2-PCF only depends on amplitude-amplitude correlations $\ensavg{\abs{\dk}\abs{\hat{\delta}(\q)}}$, while the $n$-PCFs ($n\geq3$) depend on amplitude-amplitude correlations $\ensavg{\abs{\dk}\abs{\hat{\delta}(\q)}}$, amplitude-phase correlations $\ensavg{\abs{\dk}\hat{\epsilon}(\q)}$, and phase-phase correlations $\ensavg{\ek\hat{\epsilon}(\q)}$. It follows that $\hr$ is independent of $\xi_2(r)$ and $p(k)$, up to statistical dependencies between the amplitudes $\abs{\dk}$ and phases $\hat{\epsilon}(\q)$.

\subsubsection{$\hr$ -- a parameter-free measure of cosmic filaments}

In Section \ref{subsection_physical_properties} we have established that $\hr$ is, in a limited sense, a measure of aspherical structure, such as straight filaments. This measure is statistical in that it cannot identify the individual filaments. Hence typical applications of $\hr$ are studies of cosmic structure rather than investigations of galaxies in particular environments. For the latter, sophisticated `filament-finders' have been developed, i.e., algorithms able to convert 3D galaxy distributions into skeletons of filaments \citep[e.g.][]{Bond2010,Sousbie2011}. On the downside, the relationship between those algorithms and robust measures of cosmic structure, such as correlation functions, is unknown and/or complex \citep{Pogosyan2009}. Moreover, filament-finders always require free parameters, such as user-defined thresholds and scales. By contrast, the definition of $\hr$ is parameter-free and scale-invariant; it is even independent of the numerical grid, if the cell-size is much smaller than $r$. The relationship between filament-finders and $\hr$ is therefore analogous to that between group-finders and $\xi_2(r)$: the former allow an explicit identification of individual structural components, while the latter represent mathematically robust measures of spatial statistics.

\subsubsection{$\hr$ -- a thermometer for dark matter}\label{subsubsection_thermometer}

By calculating $\hr$ for simulated cosmic density fields (Section \ref{section_cosmology}) we demonstrated that $\hr$ is more sensitive to variations of $\mdm$ than the 2-PCF $\xi_2(r)$. Therefore, measurements of $\mdm$ are significantly more accurate when based on $\hr$ than based on $\xi_2(r)$ (e.g.~Fig.~\ref{fig_mass_probability}). Moreover, $\hr$ depends on $\mdm$ out to scales at least five times larger than $\xi_2(r)$. This result is pivotal since smaller scales are those more affected by uncertain baryon physics \citep{Kim2012} masking the footprint of dark matter properties.

\subsection{Prospects of using real data}

This work paves the way towards an enhanced analysis of existing an future redshift surveys, for example to better constrain the particle mass of dark matter $\mdm$.

Any comparison between simulated and observed LSS is challenged by differences between visible matter and underlying dark matter. These effects are rather small on scales $>1\rm~\hmpc$. However, measurements on smaller scales require a precise reconstruction of the actual dark matter density field, for example using weak-lensing data, and/or a modeling of visible LSS, for example using mock-skies based on semi-analytic modeling \citep[e.g.][]{Blaizot2005,Obreschkow2009f}.

Moreover, real surveys do not come in the shape of a cubic box, but in a truncated survey volume with varying selection criteria across the volume. To deal with such masked data, our idealized formulation of the line-correlation function will need to be transcribed into a form applicable to a generic survey-mask. To do so, one might apply an approach similar to that of \cite{Landy1993}, which essentially consists in comparing the correlations in the observed density field against those in a random field with an identical survey-mask.

Finally, observational data is subject to redshift-space distortions \citep{Kaiser1987}, leading to elongated structures along the line-of-sight in the reconstructed 3D density field (`fingers-of-God'). These spurious prolate features add to the line-correlation $\hr$. This effect will require additional modeling and/or an evaluation of $\hr$ separately along radial and transverse directions, as typically done for the 2-PCF \citep{Chuang2012}. On scales larger than the redshift distortion scale, the radial and transverse parts of $\hr$ might also be used to constrain dark energy in a way analogous to the classical Alcock-Paczynski test \citep{Alcock1979}.

\subsection{Closing words}

Above all, this work demonstrates the enormous potential of phase-information. Further investigations of this information may unveil a wealth of applications, extending far beyond the case of CDM versus WDM. In light of future redshift surveys, the time seems ripe for phase-information to become a standard tool in observational cosmology. Ultimately, this field would tremendously benefit from a complete estimator of cosmic structure, i.e., a function $\mathcal{F}(\dr)$ that exclusively and exhaustively describes the information contained in a statistically homogeneous and isotropic density field $\dr$.

\section*{Acknowledgements}

Part of the research presented in this paper was undertaken as part of the Survey Simulation Pipeline (SSimPL; {\texttt{http://www.astronomy.swin.edu.au/SSimPL/}). D.O.\ was supported by the Research Collaboration Award 12105012 of the University of Western Australia. C.B.\ is supported by the Herchel Smith fund and by KingÕs College Cambridge. We thank the anonymous referee for a very constructive report.

\appendix

\section{A. Generalized convolution theorem}\label{appendix_convolution}

We here define the Fourier transform (FT) and the corresponding inverse Fourier transform (IFT) as in \cite{Peacock1999} apart from sign of $i$, 
\begin{subequations}
	\begin{align}
		\dk & = {\rm FT(\delta)} = \frac{1}{V}\int\d^Dr~e^{-i\k\cdot\r}~\dr, \label{eq_def_ft} \\
		\dr & = {\rm IFT(\hat{\delta})} = \frac{V}{(2\pi)^D}\int\d^Dk~e^{i\k\cdot\r}~\dk, \label{eq_def_ift}
	\end{align}
\end{subequations}
where $\k\in\R$ is the wavevector. Note that the reality of $\dr$ implies $\hat{\delta}(-\k)={\hat{\delta}}^\ast(\k)$, where the asterisk is the complex conjugate. Substituting $\delta$ for eq.~(\ref{eq_def_ift}) in eq.~(\ref{eq_def_Xin}), we find
\be
	\Xi_n(\r_1,...,\r_{n-1}) = \frac{1}{V}\int\d^3t\prod_{j=1}^n\left[\frac{V}{(2\pi)^3}\int_\R\d^3k_j~e^{i\k_j\cdot(\t+\r_j)}~\hat{\delta}(\k_j)\right].
\ee
Rearranging the terms,
\be
	\Xi_n(\r_1,...,\r_{n-1}) = \frac{V^{n-1}}{(2\pi)^{3n}}~\left[\prod_{j=1}^n\int_\R\d^3k_j~e^{i\k_j\cdot\r_j}~\hat{\delta}(\k_j)\right]\int_\R\d^3t~e^{i\sum_{j=1}^{n}\k_j\cdot\t}.
\ee
Solving the integral over $\t$ gives
\be
	\Xi_n(\r_1,...,\r_{n-1}) = \left[\frac{V}{(2\pi)^3}\right]^{n-1}~\left[\prod_{j=1}^{n}\int\d^3k_j~e^{i\k_j\cdot\r_j}~\hat{\delta}(\k_j)\right]\delta^3\bigg(\sum_{j=1}^{n}\k_j\bigg),
\ee
where $\delta^3$ is Dirac's delta distribution in 3D. Finally, remembering that $\r_n\equiv0$,
\be
	\Xi_n(\r_1,...,\r_{n-1}) = \left[\frac{V}{(2\pi)^3}\right]^{n-1}~\left[\prod_{j=1}^{n-1}\int\d^3k_j~e^{i\k_j\cdot\r_j}~\hat{\delta}(\k_j)\right]\hat{\delta}\bigg(-\sum_{j=1}^{n-1}\k_j\bigg),
\ee
which readily reduces to \eq{eq_connection}.

\section{B. Explicit expressions for $\xi_2$ and $\xi_3$}\label{appendix_explicit_expressions}

This paragraph explicits the Fourier space expressions of two isotropic correlation functions, which will be used in the rest of this work. First, to find the Fourier equivalent of $\xi_2(r)$, we substitute $\Xi_2(\r)$ in \eq{eq_def_xi2} for \eq{eq_2pt_convolution}, which implies
\be\label{eq_xi2_first}
	\xi_2(r) = \frac{V}{(2\pi)^D}\int\d^Dk~w_D(kr)~P(\k),
\ee
where $w_D(kr)\equiv\avg{e^{i\k\cdot\r}}_{\abs{\r}=r}$ is a weighting function. A quick calculation expanded in Appendix \ref{appendix_rot_average} shows that
\be\label{eq_wx}
	w_D(x) = \begin{cases}
		J_0(x), & \mbox{if } D=2, \\
		\sin(x)/x, & \mbox{if } D=3,
	\end{cases}
\ee
where $J_0(x)$ is the 0-th order Bessel function. In \eq{eq_xi2_first}, $w(kr)$ only depends on the integration variable $k=\abs{\k}$. The remaining $D-1$ integration variables only act on $P(\k)$. 
Performing this integration of $P(\k)$ leads to
\be\label{eq_xi2_iso}
	\xi_2(r) = \frac{V}{(2\pi)^D}\int_0^\infty\d k~S_D(k)~w_D(kr)~p(k),
\ee
where $p(k)\equiv p_2(k)=\avg{\abs{\dk}^2}_{\abs{\k}=k}$ is the isotropic power-spectrum and $S_D(k)$ denotes the surface area of the $D$-sphere, i.e.,
\be\label{eq_S}
	S_D(k) = \begin{cases}
		2\pi k, & \mbox{if } D=2, \\
		4\pi k^2, & \mbox{if } D=3.
	\end{cases}
\ee

Next, we consider the particular isotropic 3-PCF $\xi_3(r)$ for three equidistant points on a straight line, i.e.,
\be
	\xi_3(r) \equiv \avg{\Xi_3(\r,-\r)}_{\abs{\r}=r}.
\ee
After substituting $\Xi_3$ for \eq{eq_3pt_convolution}, a derivation analogous to that in Appendix \ref{appendix_rot_average} then leads to
\be\label{eq_xi3_first}
	\xi_3(r) = \frac{V^2}{(2\pi)^{2D}}\!\iint\!\d^{D\!}k\,\d^{D\!}q\,w(|\k-\q|r)~B(\k,\q).
\ee
Note that $|\k-\q|=\sqrt{k^2+q^2-kq\cos\theta}$, where $\theta\equiv\angle(\k,\q)$, only depends on the three coordinates $(k,q,\theta)$. Thus, we can first integrate $B(\k,\q)$ over the remaining $2D-3$ coordinates, which leads to
\be\label{eq_xi3_iso}
	\begin{split}
		\xi_3(r) = \,& \frac{V^2}{(2\pi)^{2D}}\!\int_0^\infty\!\!\!\!\!\!\d k\,S_D(k)\!\int_0^\infty\!\!\!\!\!\!\d q\,S_D(q)\!\int_0^\pi\!\!\!\!\d\theta\,j_D(\theta) \\
		 & \times w_D(\sqrt{k^2+q^2-2kq\cos\theta\,}~r)~b(k,q,\theta),
	\end{split}
\ee
where $b(k,q,\theta)\equiv\avg{B(\k,\q)}_{\abs{\k}=k,\abs{\q}=q,\angle(\k,\q)=\theta}$ is the isotropic bi-spectrum and $j_D(\theta)$ is the Jacobian
\be\label{eq_j}
	j_D(\theta) = \begin{cases}
		1/\pi, & \mbox{if } D=2, \\
		\sin(\theta)/2, & \mbox{if } D=3.
	\end{cases}
\ee

\section{C. Rotational average of $\exp($\lowercase{\textsl{i\,}}$\k\cdot\r)$}\label{appendix_rot_average}

In two dimensions, $\r$ is expressed in polar coordinates $r=|\r|$ and $\theta$, where $\theta$ is the angle between $\k$ and $\r$, such that $\k\cdot\r=kr\cos(\theta)$. Then,
\be
	\avg{e^{i\k\cdot\r}}_{|\r|=r} = \frac{1}{2\pi r}\int_0^{2\pi}\d\theta~r~e^{ikr\cos\theta}.
\ee
By symmetry, the real part of the integral is equal to twice the integral from $0$ to $\pi$, and, by anti-symmetry, the imaginary part of the integral vanishes, 
\be
	\avg{e^{i\k\cdot\r}}_{|\r|=r} = \frac{1}{\pi}\int_0^{\pi}\d\theta~\cos(kr\cos\theta) = J_0(kr).
\ee

In three dimensions, $\r$ is expressed in spherical coordinates $r$, $\varphi$, and $\theta$, where $\theta$ is the angle between $\k$ and $\r$, such that $\k\cdot\r=kr\cos(\theta)$. Then,
\be
	\avg{e^{i\k\cdot\r}}_{|\r|=r} = \frac{1}{4\pi r^2}\int_{0}^{2\pi}\d\varphi\int_0^{\pi}\d\theta~r^2\sin{\theta}~e^{ikr\cos\theta} = \frac{1}{2}\int_0^{\pi}\d\theta~\sin{\theta}~e^{ikr\cos\theta}.
\ee
By anti-symmetry, the imaginary part of the integral vanishes and
\be
	\avg{e^{i\k\cdot\r}}_{|\r|=r} = \frac{1}{2}\int_0^{\pi}\d\theta~\sin{\theta}~\cos(kr\cos\theta) = \frac{\sin(kr)}{kr}.
\ee

\section{D. Numerical discretization}\label{appendix_discretization}

For computational purposes, we adopt the standard numerical model: (i) the universe is described in a finite cubic box $\Omega\subset\R$ of side-length $L$ and volume $V=|\Omega|=L^D$; (ii) this box satisfies periodic boundary conditions; (iii) the density perturbation field $\dr$ is represented on a regular Cartesian grid of $N^3$ cubic cells, such that the cells have side-lengths $\Delta r=L/N$ and volumes $\Delta V=(L/N)^D$. This model is valid as long as we consider correlations on scales larger than $\Delta r$ and significantly smaller than $L$. The corresponding Fourier space discretization follows directly from the periodicity condition, which states that each mode $\k=(k_1,...,k_D)$ must satisfy $k_j L\in2\pi\mathds{N_0}~\forall j$. Therefore, the Fourier cell spacing equals $\Delta k=2\pi/L$. The side-length of the Fourier box hence becomes $N\Delta k=2\pi N/L$. This numerical discretization is illustrated in Fig.~\ref{fig_discretization} in two dimensions ($D=2$).

\begin{figure}
	\begin{center}
	\includegraphics[width=9.5cm]{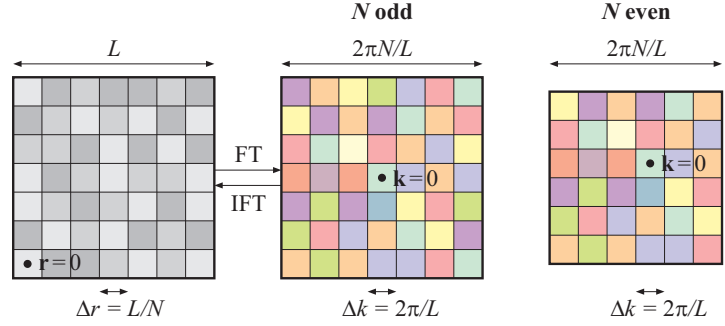}
	\caption{(Color online) Discretization rules in direct (left-most panel) and Fourier (middle and right panel) space. Complex numbers are shown in colors as described in Fig.~\ref{fig_method}. Black dots denote our choice of the origins of the coordinate systems.}
	\label{fig_discretization}
	\end{center}
\end{figure}

We are free to choose the origins, both in direct and in Fourier space. In our convention, shown in Fig.~\ref{fig_discretization}, the sets of discrete vectors $\r$ and $\k$ become
\begin{subequations}
	\begin{align}
		& \r = \Delta r\,\a \text{, where~}\a=(a_1,...,a_D) \text{~with~} a_j\in\{0,1,...,(N-1)\}, \label{eq_r_set} \\
		& \k = \Delta k\,\b \text{, where~}\b=(b_1,...,b_D) \text{~with~}b_j\in\{-\rm{floor}(N/2),...,-1,0,1,...\rm{floor}(N/2-1/2)\},  \label{eq_k_set}
	\end{align}
\end{subequations}
where ${\rm floor}(x)$ is defined as the largest integer less than or equal to $x$.

The rules for the mapping between continuous integrals and discrete sums, both in direct and Fourier space, follow directly from the expressions for $\Delta r$ and $\Delta k$. They read
\begin{subequations}
	\begin{align}
		\int\d^Dr~f(\r)~~~ & \longleftrightarrow ~~~\sum_{\r}~\Delta r~f(\r) = \frac{L^D}{N^D}\sum_{\r}~f(\r), \label{eq_mapping_direct} \\
		\int\d^Dk~\hat{f}(\k)~~~ & \longleftrightarrow ~~~\sum_{\k}~\Delta k~\hat{f}(\k) = \frac{(2\pi)^D}{L^D}\sum_{\k}~\hat{f}(\k).\, \label{eq_mapping_fourier}
	\end{align}
\end{subequations}
where the values of the functions $f(\r)$ and $\hat{f}(\k)$ on the right-hand side are cell averages. Using the mapping rules of eqs.~(\ref{eq_mapping_direct}) and (\ref{eq_mapping_fourier}), the FT and IFT of eqs.~(\ref{eq_def_ft}) and (\ref{eq_def_ift}) become the discrete FT (DFT) and the inverse DFT (IDFT), respectively,
\begin{subequations}
	\begin{align}
		\dk & = \frac{1}{N^D}\sum_{\r}e^{-i\k\cdot\r}~\dr, \label{eq_def_ft_discr} \\
		\dr & = \sum_{\k}e^{i\k\cdot\r}~\dk. \label{eq_def_ift_discr}
	\end{align}
\end{subequations}
By virtue of the same rules, $\xi_2(r)$ in \eq{eq_xi2_first}, $\xi_3(r)$ in \eq{eq_xi3_first}, and $\hr$ in \eq{eq_def_linecorrelation} become
\begin{subequations}\label{eq_correlations_discretized}
	\begin{align}
		\xi_2(r) & = \sum_{\k} w_D(kr)~P(\k), \\
		\xi_3(r) & = \sum_{\k}\sum_{\q}w_D(|\k-\q|r)~B(\k,\q), \\
		\h(r) & = \left(\frac{r}{L}\right)^{\frac{3D}{2}}\!\!\sum_{\abs{\k}\leq\pi/r}\sum_{\abs{\q}\leq\pi/r}w_D(|\k-\q|r)\,\frac{B(\k,\q)}{\Abs{B(\k,\q)}}.
	\end{align}
\end{subequations}
These are the three functions, which we calculated in the examples, e.g.~Fig.~\ref{fig_examples}.
 

\end{document}